\documentclass[prl,twocolumn,showpacs]{revtex4}

\usepackage{bm}
\usepackage{graphicx}
\usepackage{times}

\newcommand*{\be}{\begin{equation}}
\newcommand*{\ee}{\end{equation}}
\newcommand{\ba}{\begin{eqnarray}}
\newcommand{\ea}{\end{eqnarray}}

\begin{document}

\title{Gauge Theory for the Rate Equations: Electrodynamics on a Network}

\author{Carsten Timm}
\email{ctimm@ku.edu}
\affiliation{Department of Physics and Astronomy, University of Kansas,
Lawrence, KS 66045, USA}

\date{December 22, 2006}

\begin{abstract}Systems of coupled rate equations are ubiquitous in many areas
of science, for example in the description of electronic transport through
quantum dots and molecules. They can be understood as a continuity equation
expressing the conservation of probability. It is shown that this conservation
law can be implemented by constructing a gauge theory akin to classical
electrodynamics on the network of possible states described by the rate
equations. The properties of this gauge theory are analyzed. It turns out that
the network is maximally connected with respect to the electromagnetic fields
even if the allowed transitions form a sparse network. It is found that the
numbers of degrees of freedom of the electric and magnetic fields are equal.
The results shed light on the structure of classical abelian gauge theory
beyond the particular motivation in terms of rate equations.\end{abstract}


\maketitle

\textit{Introduction}.---Let
us consider a system that assumes states $|i\rangle$ with probabilities
$P_i$, where the states form a finite or countable set. The
probabilities must add up to unity, $\sum_i P_i = 1$. If the rates
of change of the probabilities are {linear} functions of the
probabilities, we can write
\be
\dot P_i = \sum_j ( R_{ij}P_j - R_{ji}P_i ) .
\label{1.rate}
\ee
This is a set of {rate equations}, containing the transition rates $R_{ij}$
from state $|j\rangle$ to state $|i\rangle$. The first term under the sum
describes transitions from other states to $|i\rangle$, whereas the second
describes transitons out of state $|i\rangle$.
Often, many of the $R_{ij}$ are zero. Importantly, Eq.~(\ref{1.rate})
conserves probability: $\sum_i \dot P_i = 0$.


Rate equations are ubiquitous in science, in particular in
physics, chemistry, and biology. They describe systems far
from equilibrium, such as lasers \cite{Sie86}, semiconductor devices
\cite{Car02}, quantum dots \cite{GlM88}, chemical reactions \cite{Con90},
enzyme kinetics \cite{Pur96}, and biological populations \cite{Hop82}.


As a specific example, we discuss
a molecular quantum dot coupled to leads. The time evolution of the
{complete} system of molecule and leads is described by the von Neumann
equation for the full density operator, $\dot\rho = -i[H,\rho]$, where $H$ is
the Hamiltonian of the complete system and $\hbar=1$. If one is interested
in the properties of the molecule, it is advantageous to integrate out the
states of the leads to obtain an equation of motion (master equation) for the
{reduced} density matrix $\rho^d$ in the many-particle Hilbert space of
the molecule alone
\cite{Nak58,Zwa60,ToM75,Blu81,Ahn94,ScS94,KSS95,TuM02,BrF04,Mitra,Elste,Koch}.
This requires approximations. For example, if the tunneling amplitude between
molecule and leads is small compared to the typical molecular energy-level
spacing, one can apply perturbation theory
\cite{Blu81,ScS94,KSS95,TuM02,BrF04,Mitra,Elste,Koch}. The master equation
contains diagonal and off-diagonal components of $\rho^d$. If one assumes that
the off-diagonal components, i.e., the superpositions, decay rapidly
on the relevant timescale \cite{Zur82}, one is left with coupled rate equations
(\ref{1.rate}) for the diagonal components, which are just the probabilities
$P_i \equiv \rho^d_{ii}$ of molecular many-body states.

Our discussion is not specific to a particular
incarnation of the rate equations. Rather, it starts from the general principle
of conservation of probability. We consider the \emph{network} formed by
the states $|i\rangle$. The transitions between
them are the edges of this network.
The rate equations (\ref{1.rate}) can be written
as a \emph{continuity equation} on the network \cite{ZiS06},
\be
\dot P_i = \sum_j J_{ij} ,
\label{1.cont}
\ee
where $J_{ij}$ is a {probability current} defined as an
{antisymmetric} quantity on the edges \cite{footnote.antisym},
\be
J_{ij} \equiv R_{ij}P_j - R_{ji}P_i .
\label{1.Jdef}
\ee
The conservation law (\ref{1.cont}) suggests to look for a gauge theory that
implements it \cite{Zia06}. The conserved {scalar} field
suggests a generalization of electrodynamics.


There is a sizeable literature on electrodynamics 
\cite{Yee66,Che94,TeC99,HeT05} and non-abelian gauge theories
\cite{Wil74,Smi02,Rot05} on {regular} lattices, mainly motivated by
discretizing continuum theories to facilitate numerical calculations. Chew
\cite{Che94} introduces a discrete vector calculus to formulate electrodynamics
on a lattice. A Lagrangian approach has been used to obtain a network
approximation for electrodynamics in certain electronic devices \cite{ArP04}.
An electronic-network model for coupled linear chemical reactions has also been
proposed \cite{Shi84}. In this model, concentrations and not probabilities are
mapped onto charges and a magnetic field or gauge fields are not introduced.


The present Letter can also be read as a generalization of classical
electrodynamics to a network and is thus of interest for the fundamental
understanding of gauge theories, independently of the motivation from rate
equations. It is not obvious how one should generalize Chew's discrete vector
calculus \cite{Che94} to a network, since there are no natural definitions of
forward and backward differences. The Maxwell equations in {integral} form
are more easily generalized. For example, the electric flux through a ``wall''
that divides the network into two parts is a meaningful quantity. It turns out
that due to the lack of a length scale the electric (magnetic) field is the
same as the electric (magnetic) flux and there is no difference between the
differential and integral forms of the Maxwell equations.


\textit{Maxwell equations}.---Our goal is to
construct electric and magnetic fields that
implement the conservation of probability.
If we view $P_i$ as a charge density, it should provide the sources of the
electric field. We define an antisymmetric field $E_{ij}$ on the edges.
Then Gauss' Law should read
\be
\sum_j E_{ji} = P_i .
\label{1.GaussE1}
\ee
The sum over all outgoing electric fields equals the enclosed charge.
Note that $E_{ji}$ is generally nonzero also if transitions from $|i\rangle$ to
$|j\rangle$ are forbidden so that $J_{ji}=0$.
This is not surprising since in standard electrodynamics there is certainly an
electric field in an insulator.
As far as the electric (and, as we will see, the magnetic) field is
concerned, the network is maximally connected, every node is the neighbor of
every other node.


Equation (\ref{1.GaussE1}) leads to a problem: Summing over $i$ we find
\be
0 = \sum_{ij} E_{ji} = \sum_i P_i \stackrel{?}{=} 1 .
\label{1.GaussE2}
\ee
We can rectify this by adding an additional fictitious node $0$ with $P_0 = -1$
and $R_{i0}=R_{0i}=0$ for all $i$. Node $0$ does not affect the rate equations
for the real states but makes the network charge neutral.
The problem is that in Eq.~(\ref{1.GaussE2}) we obtain a relation between
the flux through the surface of the system
and its total charge. However, our network does not have a surface. The same
problem arises if one tries to construct electrodynamics on any compact
space.

\begin{figure}[ht]
(a)\includegraphics[width=1.5in]{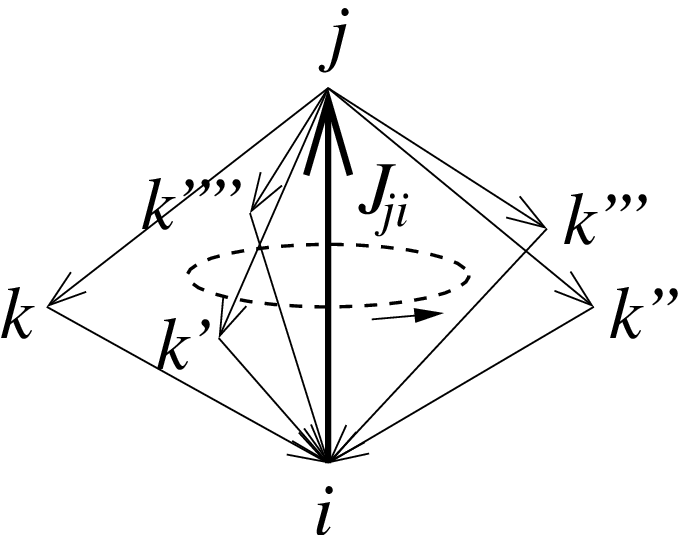}
(b)\includegraphics[width=1.5in]{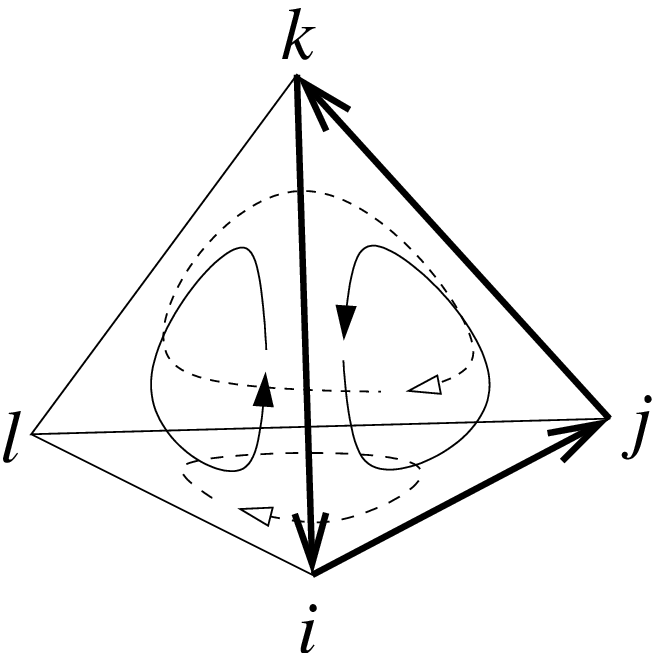}
\caption{\label{fig.AMlaw}(a) Sketch showing that
the Amp\`ere-Maxwell Law relates the current
on the edge from $i$ to $j$ to the circulation of the
magnetic field around this edge.
(b) Sketch of an elementary cell on the network,
bounded by four plaquettes oriented outwards.}
\end{figure}

Next, we define a magnetic field $B_{(ijk)}$ on the elementary oriented
{plaquettes}
of the network, which are oriented triangles $(ijk)$. The magnetic field is
invariant under cyclic commutation of $i,j,k$ and changes sign for
anti-cyclic commutations. To
have $J_{ij}$ generate the magnetic field, we write
the Amp\`ere-Maxwell Law as
\be
\sum_k B_{(ijk)} - \frac{1}{c}\,\dot E_{ji} = \frac{1}{c}\,J_{ji}  .
\label{1.AM1}
\ee
Figure \ref{fig.AMlaw}(a) shows
that the equation contains a sum over the magnetic field penetrating all
plaquettes adjacent to the edge carrying the current, corresponding to a line
integral. $E_{ji}$ and $B_{(ijk)}$ have the
same dimension. Then $c$ is a frequency, not a velocity, since the network does
not have a length scale.


Faraday's Law should relate the time derivative of the magnetic field
through a plaquette to the sum over electric fields along its edges.
This is achieved by
\be
E_{ji} + E_{kj} + E_{ik} + \frac{1}{c}\,\dot B_{(ijk)} = 0 .
\label{1.Far1}
\ee
The sum $E_{ji} + E_{kj} + E_{ik}$ is the analogue
of the electromotive force around the plaquette so that its
inductance is $1/c$.

Finally, the magnetic flux through the surface of any volume should vanish. An
elementary cell on the network can be characterized by four nodes $i,j,k,l$.
Then,
\be
B_{(ijk)} + B_{(jlk)} + B_{(lik)} + B_{(ilj)} = 0 .
\label{1.GaussB1}
\ee
Figure \ref{fig.AMlaw}(b) shows that the orientation of all
faces is the same.



The Maxwell equations imply the continuity equation: Summation of
Eq.~(\ref{1.AM1}) over $j$ gives
\be
\sum_{jk} B_{(ijk)} - \frac{1}{c}\sum_j \dot E_{ji} = \frac{1}{c}\sum_j J_{ji} .
\ee
The first term vanishes by symmetry. Using Eq.~(\ref{1.GaussE1}) we reobtain
Eq.~(\ref{1.cont}).
Note that this derivation also works if the rates $R_{ij}$ depend on time.
In fact, the derivation does not make any assumption about the
dependence of $J_{ij}$ on $P_i$.

\textit{Coulomb and Biot-Savart Laws}.---We
now consider the solution of the Maxwell equations for the fields. Direct
calculation yields $P_i - P_j = NE_{ji} + c^{-1}\sum_k \dot B_{(ijk)}
= NE_{ji} + c^{-2}\ddot E_{ji} + c^{-2}\dot J_{ji}$.
Here, $N$ is the number of nodes, including node $0$.
In the static case, we thus find the {Coulomb Law}
\be
E_{ji} = \frac{1}{N}\,(P_i - P_j) .
\label{1.Coul3}
\ee
Interestingly, it is
local: The static electric field on an edge is completely determined by the
charges on the adjacent nodes.

We also find
$c^{-1}(J_{ji} + J_{kj} + J_{ik}) = NB_{(ijk)} + c^{-2}\ddot B_{(ijk)}$.
In the static limit we obtain a local {Biot-Savart Law}
\be
B_{(ijk)} = \frac{1}{cN}\, (J_{ji} + J_{kj} + J_{ik}) .
\label{1.BS3}
\ee

\textit{Gauge fields}.---As in standard electrodynamics, we introduce gauge
fields to satisfy the homogeneous Maxwell equations (\ref{1.Far1}) and
(\ref{1.GaussB1}). The magnetic field is written as
\be
B_{(ijk)} = A_{ji} + A_{kj} + A_{ik} ,
\label{1.A1}
\ee
where $A_{ij}$ is antisymmetric. The form on the right-hand side is
analogous to a curl,
compare Eq.~(\ref{1.Far1}). Interestingly, the magnetic field has
many more components than the ``vector potential'' $A_{ji}$, which
is defined on edges, not plaquettes.


Then Faraday's Law takes on the form
$E_{ji} + \dot A_{ji}/c + E_{kj} + \dot A_{kj}/c + E_{ik} + \dot A_{ik}/c = 0$,
which suggests to write
\be
E_{ji} = -(\phi_j - \phi_i) - \frac{1}{c}\,\dot A_{ji} ,
\ee
where the ``scalar potential'' $\phi_i$ is defined on the nodes.
It is easy to show that Eqs.~(\ref{1.Far1}) and (\ref{1.GaussB1}) are indeed
satisfied.

The electric and magnetic fields are invariant under the simultaneous gauge
transformations
\ba
A_{ji} & \to & A_{ji} + \Lambda_j - \Lambda_i ,
\label{1.gauge1}
\\
\phi_i & \to & \phi_i - \frac{1}{c}\,\dot \Lambda_i ,
\label{1.gauge2}
\ea
where $\Lambda_i$ are arbitrary time-dependent functions. One can now consider
specific gauge choices. For example, the analogue of the Lorenz gauge requires
\be
\sum_j A_{ji} + \dot\phi_i = 0 , \qquad
\sum_i \phi_i = 0 .
\label{1.Lo2}
\ee
It is easy to show that this choice is possible. Then the remaining
two Maxwell equations take on the simple form
\be
N\phi_i + \frac{1}{c^2}\,\ddot \phi_i = P_i , \qquad
NA_{ji} + \frac{1}{c^2}\,\ddot A_{ji} = \frac{1}{c}\,J_{ji} .
\ee
These equations are coupled by the gauge condition (\ref{1.Lo2}).

The representation in terms of gauge fields facilitates the discussion of
the number of dynamic degrees of freedom. For a regular lattice, He and
Teixeira \cite{HeT05} show that the electric and
magnetic fields contain the same number of independent degrees of freedom. For
the network, we first reformulate the question by asking how many
electric and magnetic field components can consistently and independently be
specified.


We choose the gauge $\phi_i=0$ so that the $A_{ij}(t)$ are $N(N-1)/2$
independent functions. Since the Maxwell equations lead to second-order
differential equations for the $A_{ij}(t)$, we can and must specify two initial
conditions for each to determine the solution. This gives $N(N-1)$
independent degrees of freedom. If we specify the initial values for the
electric and magnetic fields instead, we must specify all $N(N-1)/2$ components
of $E_{ij}$ to fix $\dot A_{ij}$. We then must also specify $N(N-1)/2$
components of $B_{(ijk)}$ to determine the solution. Thus the number of
dynamic degrees of freedom in the electric and magnetic field is the same also
on the network.

\textit{Lagrangian and energy}.---The Maxwell equations can be concisely
expresses by the Lagrangian
\be
L = \frac{1}{2} \sum_{\langle ij\rangle} E_{ij}^2
  - \frac{1}{2} \sum_{(ijk)} B_{(ijk)}^2
  - \sum_i P_i \phi_i
  + \frac{1}{c} \sum_{\langle ij\rangle} J_{ij} A_{ij} .
\label{1.L1}
\ee
As usual, $E_{ij}$ and $B_{(ijk)}$ are to be expressed in terms of the gauge
fields. The sum $\sum_{\langle ij\rangle}$ is over all edges, counting each
edge once, and $\sum_{(ijk)}$ is over all elementary
plaquettes, counting each plaquette once. The action is $\mathcal{S} =
\int dt\, L$. It is straightforward to show that the Euler-Lagrange equations
for Hamilton's principle $\delta S=0$ reproduce the two inhomogeneous Maxwell
equations.
%
Incidentally, a {covariant} formulation is not possible due to
the different structure of space (network) and time (continuum).


The Lagrangian suggests to define the energy densities  of the electric and
magnetic fields as  $w^{\mathrm{el}}_{ji} \equiv E_{ji}^2/2$ and
$w^{\mathrm{mag}}_{(ijk)} \equiv B_{(ijk)}^2/2$, respectively.
The local energy balance can then be expressed using
the ``Poynting vector'' $S_{ji}^k \equiv -c E_{ji} B_{(ijk)}$.
With this definition we obtain
\ba
\frac{d}{dt} w^{\mathrm{el}}_{ji} & = & -\sum_k S_{ji}^k - E_{ji}J_{ji} , \\
\frac{d}{dt} w^{\mathrm{mag}}_{(ijk)} & = & - S_{ji}^k
  - S_{kj}^i - S_{ik}^j .
\ea
Energy is not conserved but changes due to ohmic dissipation.
However, unlike in standard electrodynamics, the energy can actually
{increase},
since $E_{ji}J_{ji}$ can be negative for asymmetric rates $R_{ij}$.
Note that the generalization of the cross product yields a
peculiar object $S_{ji}^k$, which is symmetric only in its two lower indices.

\textit{Medium equation}.---The Maxwell equations do not yet form a closed set.
As in standard electrodynamics in media one has to complement the Maxwell
equations by {medium equations} describing the response of the medium.

The situation most closely resembling our case is that of a conductor. Ohm's Law
for the network would read $J_{ij}=\sigma_{ij}E_{ij}$. However, in our case the
medium equation is just the definition of the current, Eq.~(\ref{1.Jdef}). This
equation is conceptually different from Ohm's Law in that it expresses the current
density in terms of the charges, not the electric field. The origin is that we are
actually describing a {diffusive} system.


We \emph{can} rewrite the medium equation to resemble Ohm's Law: With
Gauss' Law (\ref{1.GaussE1}) we obtain
\be
J_{ji} = \sum_k (R_{ji} E_{ki} + R_{ij} E_{jk}) .
\ee
This is a \emph{nonlocal} version of Ohm's Law on the network, where the rates
$R_{ij}$ play the role of conductivities. However, since $R_{ij}$ is not
symmetric, nonzero currents can be present in the stationary state.

If the rates are symmetric, $R_{ij}=R_{ji}$, which is a sufficient but not
necessary condition for detailed balance \cite{ZiS06},
we can rewrite the medium equation as
\be
J_{ji} = NR_{ji}\, \bigg( E_{ji} + \frac{1}{cN} \sum_k \dot B_{(ijk)} \bigg) .
\ee
Here, $NR_{ji}$ acts as the conductivity and the expression in parentheses is
the force per unit charge. The first term is the normal electric field, while
the second is not present in standard electrodynamics. It is the rate of change
of magnetic circulation around the edge from $j$ to $i$.

\textit{Discussion and conclusions}.---We have formulated classical
electrodynamics on a network of possible states, motivated by the conservation
of probability in systems of coupled rate equations. We close with a number of
remarks.

(i) Even if the transition rates are nonzero only between certain states $i,j$,
we have to introduce electromagnetic fields on \emph{all} possible edges
$\langle ij\rangle$ and plaquettes $(ijk)$ of the network. Thus this system does
not reduce to electrodynamics on a lattice \cite{Che94} for the case that the
nodes connected by allowed transitions form a regular lattice. Nevertheless, the
number of degrees of freedom of the electric and magnetic fields on the network
are equal, as for a regular lattice.

(ii) As in standard electrodynamics, only the two inhomogeneous Maxwell equations
are required to  implement the continuity equation. Thus the theory allows to
introduce magnetic monopoles, which here live on the dual network
formed by the cells of the original network.

(iii) Compared to continuum electrodynamics, scalar fields correspond to
quantities defined at the nodes ($P_i$, $\phi_i$), polar-vector fields
correspond to quantities defined on edges ($J_{ij}$, $E_{ij}$, $A_{ij}$), and
axial-vector fields correspond to plaquette fields ($B_{(ijk)}$). A Poynting
vector can be defined, but has a more complicated structure than the ``vector''
$E_{ij}$ because electric and magnetic fields live in different places.


(iv) We have reformulated the rate equations as a variational principle,
$\delta S=0$, for the action of electromagnetic fields. What is gained by this
formulation? On the one hand, it represents a new way to think about rate
equations within the framework of electrodynamics. It works for \emph{any}
dependence of the current on the probabilities and their time derivatives, as
long as $J_{ij}$ is antisymmetric.


The framework is expected to be useful for {inverse} problems: How can one
construct a dynamical system with specified properties? As an example,
note that the new fields $E'_{ji} \equiv E_{ji} + \sum_k \dot B_{(ijk)}/cN$,
$B'_{(ijk)} \equiv 0$ and the new current $J'_{ji} \equiv J_{ji} - c \sum_k
B_{(ijk)} - \sum_k \ddot B_{(ijk)}/cN$ satisfy the Maxwell equations with the
{original} charges $P_i$. Since $B'=0$, $J'_{ji}+J'_{kj}+J'_{ik}$ vanishes for
all $i,j,k$, cf.\ the discussion leading to Eq.~(\ref{1.BS3}). This means that for
{any} time-dependent probabilities $P_i$ one can find a dynamical system with
these probabilities and with $J_{ji}+J_{kj}+J_{ik}=0$. Since the current is
curl-less, one can write it in terms of a ``current potential,'' $J_{ji} =
-(\psi_j-\psi_i)$. Gauss' Law then shows that $\psi_i=-\dot P_i/N$ is a solution.


(v) The framework is of interest beyond the motivation in
terms of rate equations, since it shows that a classical gauge theory can be
formulated consistently on a network, which does not have a natural metric. For
this reason the parameter $c$ is a frequency instead of a velocity. It is an
interesting question how non-abelian gauge theories fare in this context.

The author would like to thank R. K. P. Zia, J. M. Arnold,
F. S. Nogueira, and J. P. Ralston for illuminating discussions.


\begin{thebibliography}{99}
\bibitem{footnote.antisym}For probability to be conserved we need both
Eq.~(\protect\ref{1.cont}) and the antisymmetry of $J_{ij}$---if $J_{ij}$ is
not antisymmetric, probability is not conserved on the edges.
\end{thebibliography}
\end{document}